\documentclass[preprint]{revtex4-1}
\usepackage{graphics}
\usepackage{verbatim}
\usepackage{amsmath}
\usepackage[dvipsnames]{xcolor}
\usepackage{bm}
\newcommand{\bu}{{\boldsymbol u}}
\newcommand{\bs}{\boldsymbol}
\begin{document}
\title{Vortex dynamics and Lagrangian statistics in a model for active turbulence}
\author{Martin James} \author{Michael Wilczek} \email{michael.wilczek@ds.mpg.de}     
\affiliation{ Max Planck Institute for Dynamics and Self-Organization (MPI DS), Am Fa{\ss}berg 17, 37077 G\"{o}ttingen, Germany}
\date{February 14, 2018}
\begin{abstract}
Cellular suspensions such as dense bacterial flows exhibit a turbulence-like phase under certain conditions. We study this phenomenon of ``active turbulence'' statistically by using numerical tools.  Following Wensink et al. [Proc. Natl. Acad. Sci. U.S.A  109:14308 (2012)], we model active turbulence by means of a generalized Navier Stokes equation. Two-point velocity statistics of active turbulence, both in the Eulerian and the Lagrangian frame, is explored. We characterize the scale-dependent features of two-point statistics in this system. Furthermore, we extend this statistical study with measurements of vortex dynamics in this system. Our observations suggest that the large-scale statistics of active turbulence is close to Gaussian with sub-Gaussian tails.
\end{abstract}
\maketitle
\section{Introduction}
\label{intro}

Active systems such as a flock of birds, a swarm of bacteria or active colloids form fascinating meso-scale structures with long-range order exceeding the sizes of individual agents by an order of magnitude or more \cite{yeomans2016hydrodynamics,menzel2015tuned,soto2015self,elgeti2015physics,marchetti2013hydrodynamics}. Theories describing the formation and evolution of such meso-scale coherent structures in active systems have been a topic of active research in the past two decades \cite{marchetti2013hydrodynamics,vicsek1995novel,czirok1997spontaneously,toner1998flocks,vicsek2012collective,choudhary2015effect,ramaswamy2010mechanics,chatterjee2016front}. It is known that the core features of these diverse phenomena can be modeled by taking into account just a few dynamical effects such as self-propulsion and inter-particle interactions \cite{marchetti2013hydrodynamics,vicsek1995novel,czirok1997spontaneously}.

Arguably the most diverse of these phenomena occurs at the smallest of the biological scales where collective dynamics of microbes or intra-cellular structures results in interesting spatio-temporal patterns and non-trivial dynamical features. Among these is the phenomenon of ``active turbulence" -- chaotic dynamics of dense suspensions -- which has been observed in bacterial as well as microtubule systems \cite{dombrowski2004self,sanchez2012spontaneous}. In particular ``bacterial turbulence" has been recently observed in quasi two-dimensional suspensions of {\it B. Subtilis} \cite{wensink2012meso,wensink2012emergent}. While the phenomenon shows considerable qualitative similarity with hydrodynamic turbulence by virtue of which it gets its name, active turbulence displays an intrinsic length-scale selection absent in hydrodynamic turbulence, which is characterized by the formation of stable vortices of approximately constant sizes \cite{wensink2012meso,bratanov2015new}. This typical scale is larger than the scale on which the driving occurs as the result of an upscale energy transport. This inverse energy transfer is well-known from two-dimensional Navier-Stokes turbulence forced at small scales \cite{kraichnan1980two,boffetta2000inverse}. The main difference is that in a passive Navier-Stokes fluid, the forcing has to be applied externally.

\begin{figure*}
\resizebox{0.30\textwidth}{!}{
  \includegraphics{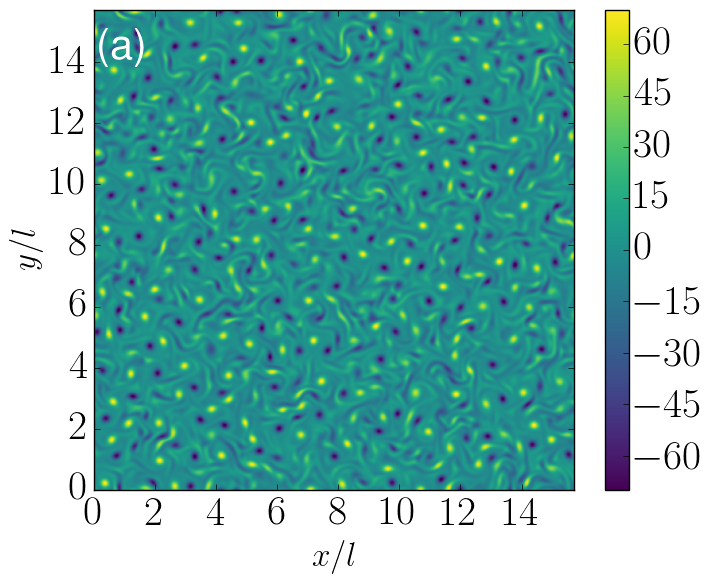}
}  
  \resizebox{0.34\textwidth}{!}{
  \includegraphics{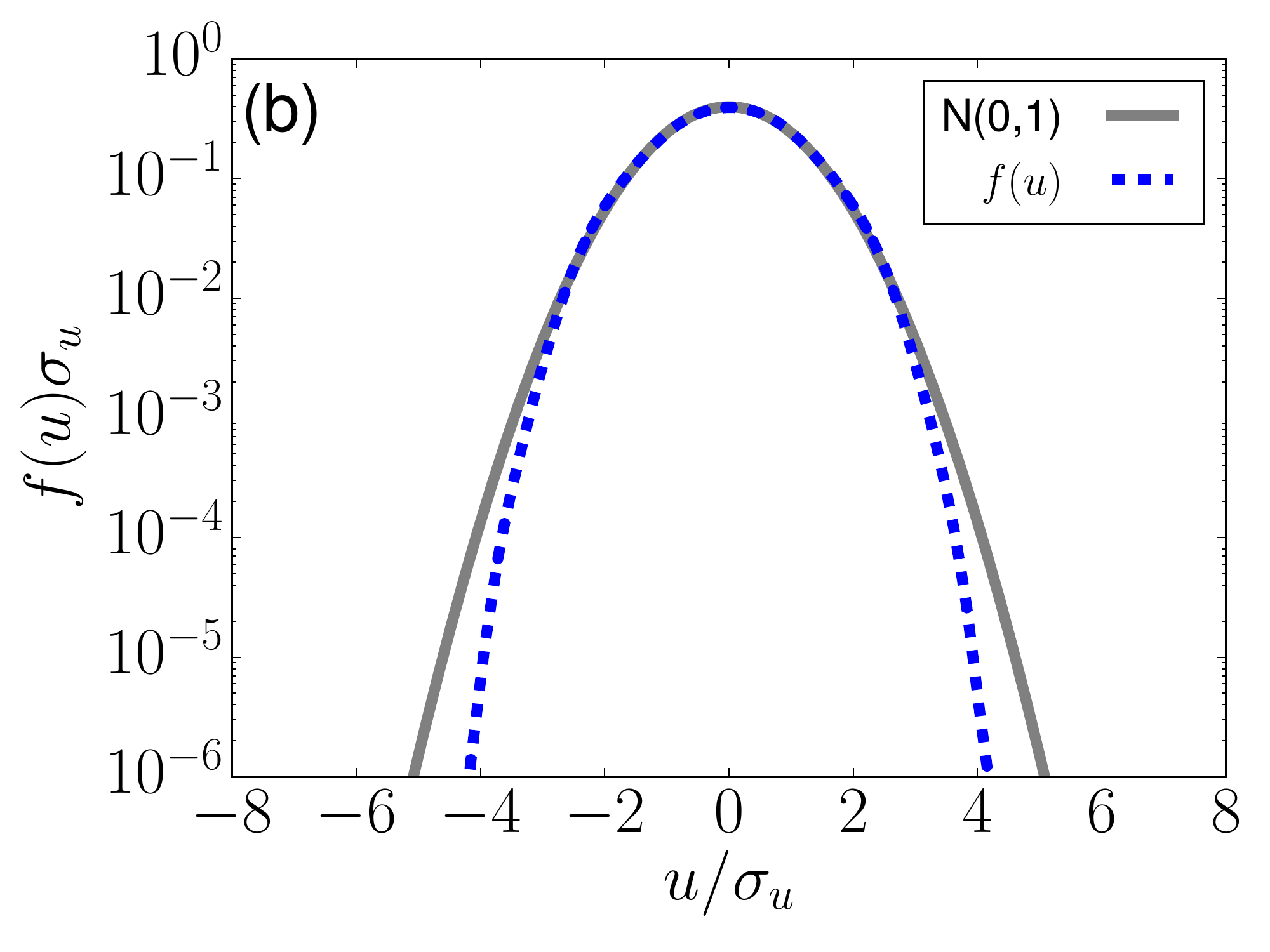}
}
\resizebox{0.34\textwidth}{!}{
  \includegraphics{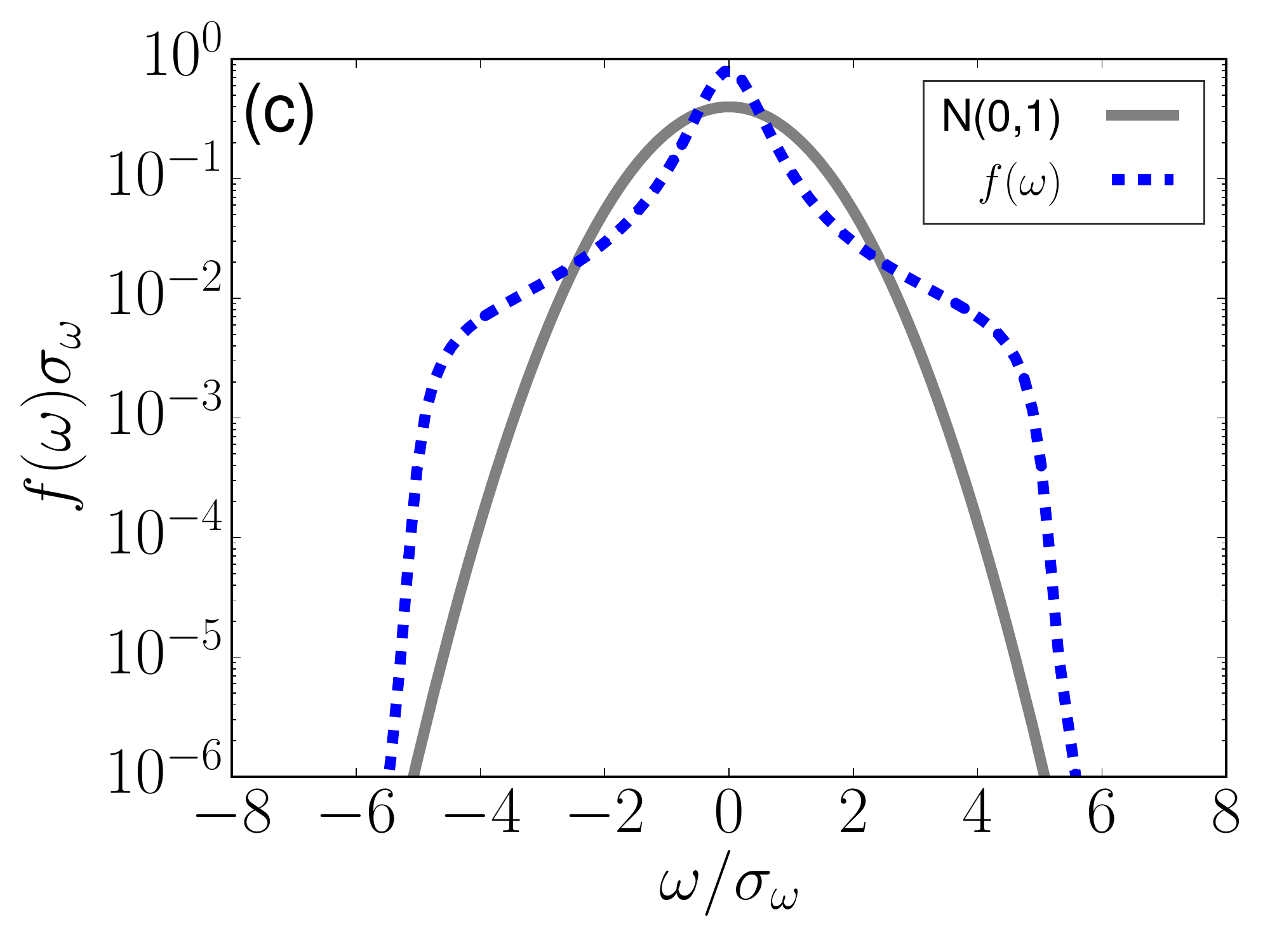}
}
\\
\resizebox{0.30\textwidth}{!}{
  \includegraphics{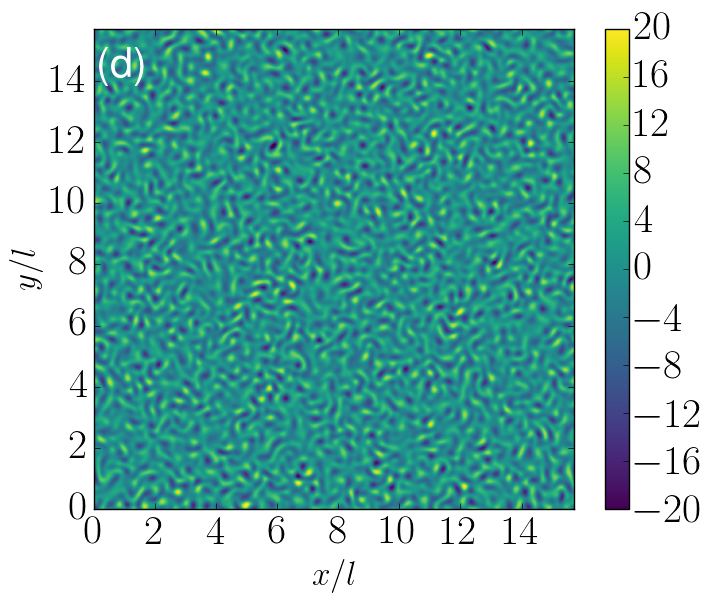}
}  
  \resizebox{0.34\textwidth}{!}{
  \includegraphics{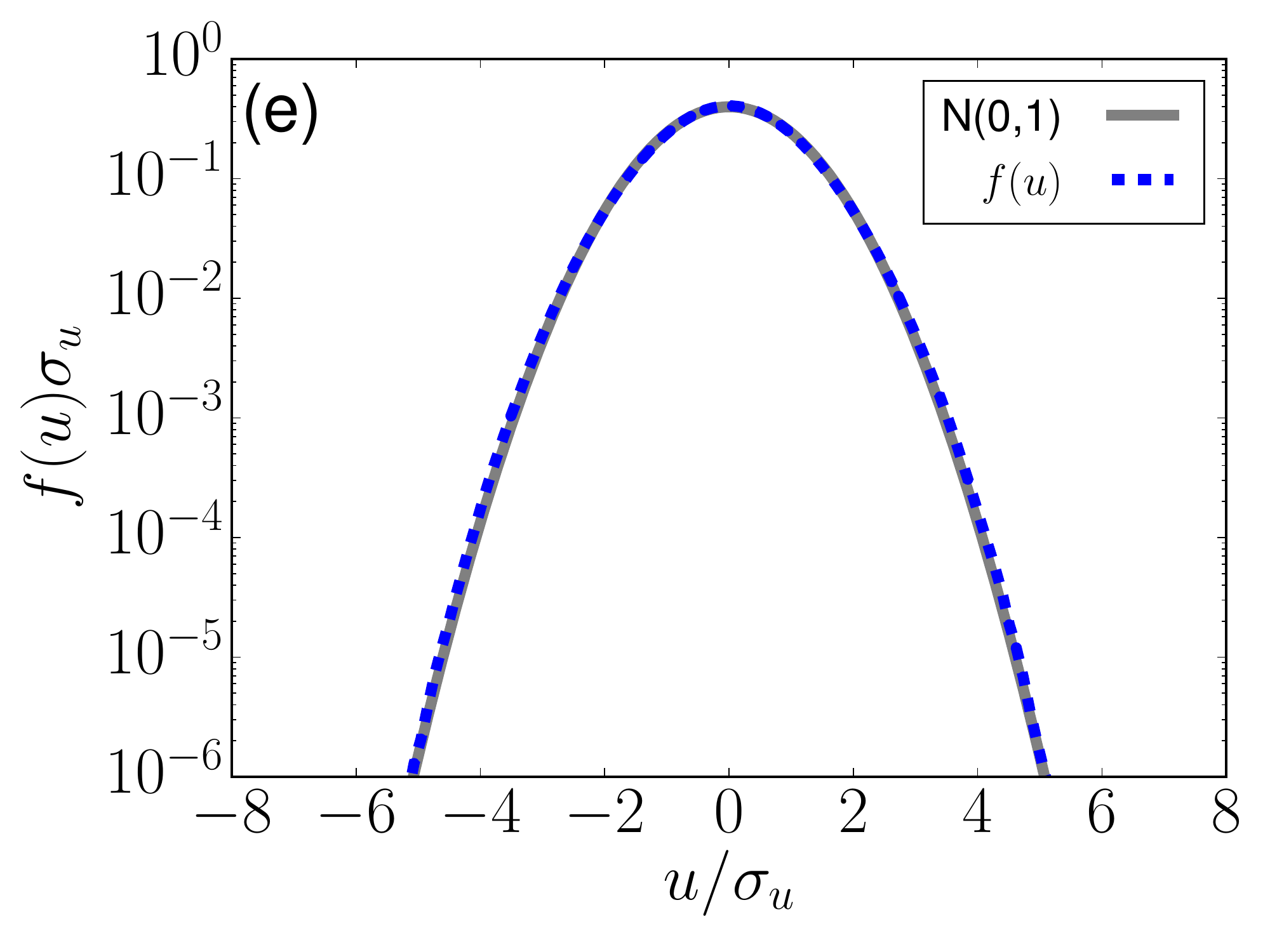}
}
\resizebox{0.34\textwidth}{!}{
  \includegraphics{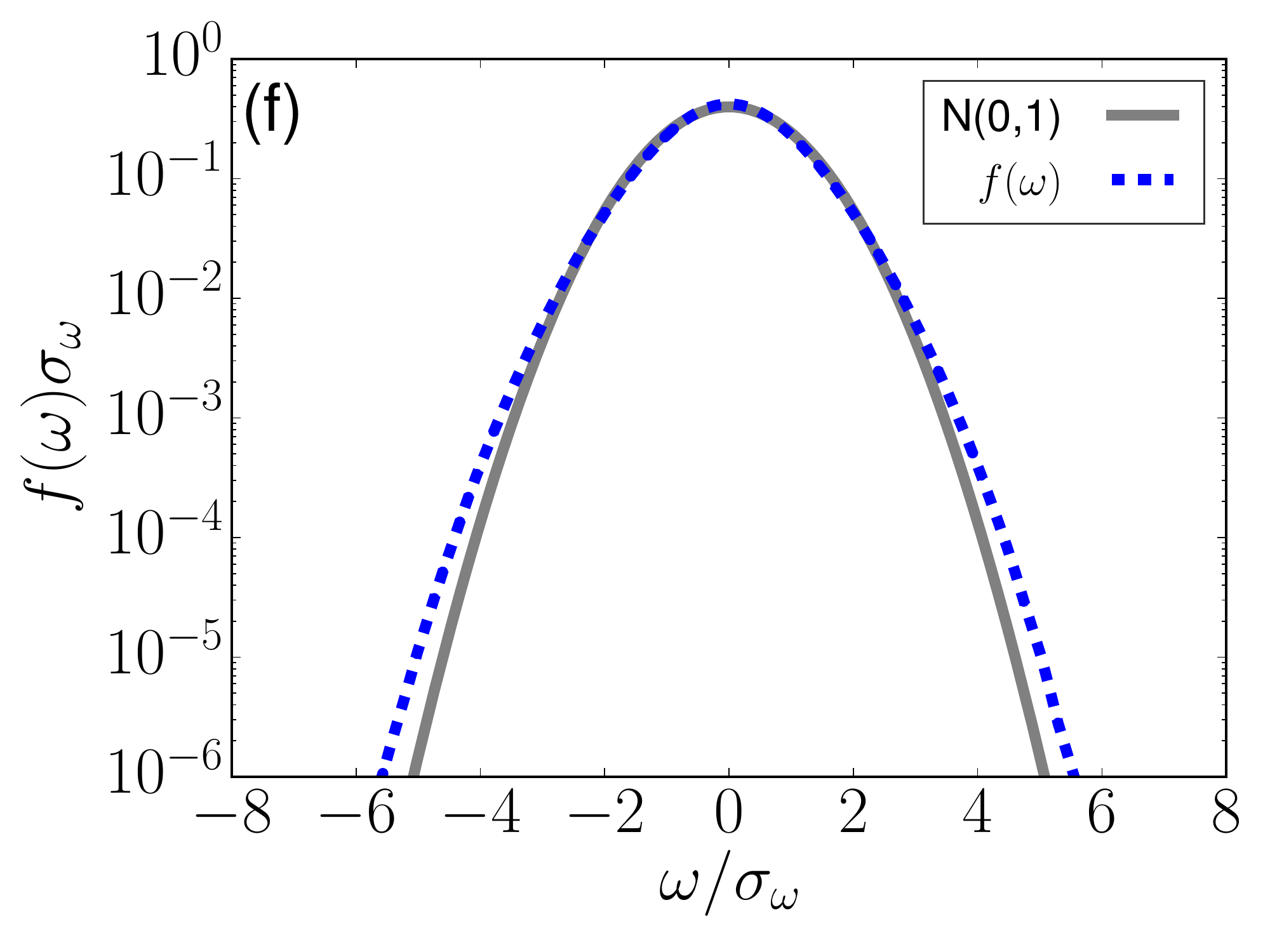}
}
\caption{Upper row: active turbulence state with broad-band forcing ($\alpha=-1$). Panel (a) shows a snapshot of the vorticity field of active turbulence obtained through direct numerical simulation of eq. \eqref{eq:active} with parameters chosen according to ref. \cite{bratanov2015new}. Note that the vortices are approximately of the same size exemplifying the selection of a length scale in this system. The single-point velocity and vorticity distributions are shown in panels (b) and (c), respectively; $\sigma_u$ and $\sigma_\omega$ are the standard deviations of the respective PDFs. The single-point velocity PDF is close to Gaussian, but has slightly sub-Gaussian tails. The vorticity PDF deviates considerably from Gaussian. Lower row: weakly excited case ($\alpha=4$). Compared to the active turbulence case, the snapshot (d) shows less pronounced vortex structures. The single-point velocity and vorticity PDFs, (e) and (f), respectively, are very close to Gaussian.}
\label{fig:eulerian}
\end{figure*}

The chaotic nature of active turbulence calls for a statistical investigation and forms the subject matter of this study. Our objective here is to provide an extensive statistical study of this phenomenon by using numerical simulations.
Our analysis is based on a recently introduced minimal continuum model for active turbulence \cite{wensink2012meso}, the details of which are presented in sect. \ref{sec:model}. As an example, fig. \ref{fig:eulerian}(a) shows a snapshot of the vorticity field of the active turbulent system in the statistically stationary state obtained through direct numerical simulation. The corresponding supplementary movie 1 shows the evolution of this field with time. Note that the intense vortices in this system are stable and have a long lifetime.

In this work, we study both the Eulerian and the Lagrangian properties as well as the characteristics of vortex dynamics in this system. Previous works on this subject have dealt with the Eulerian properties of the active turbulence field, see {\it e.g.} ref. \cite{bratanov2015new}. In sect. \ref{sec:euler} we extend this further with two-point velocity statistics and vorticity statistics to set a reference for the subsequent investigations after introducing the active turbulence model in sect. \ref{sec:model}. In sect. \ref{sec:lagrangian} we study the transport properties of the active turbulence field by investigating both the vortex dynamics and the Lagrangian features.

In the context of active turbulence, Lagrangian measurements describe the properties of tracer particles of the locally averaged velocity of the bacterial field, providing insights into transport properties and  mixing of bacterial suspensions. Measurements of this kind provide a framework to better understand the experimental works on bacterial dispersion \cite{huang2017taylor} and dynamics of small objects in bacterial baths \cite{wu2000particle}.

\section{The active turbulence model}
\label{sec:model}

Regarding the mathematical modeling of active flows, a continuum description appears suitable whenever larger-scale flow structures compared to the individual extents of the active agents are of interest. For example, such a continuum description has been established based on a coupled set of equations of two order parameter fields -- the velocity field and the local orientation of the active agents \cite{thampi2016active,giomi2015geometry,thampi2014instabilities,urzay2017multi}. This level of description is particularly useful for characterizing the role of defects on the active dynamics \cite{giomi2014defect,elgeti2011defect}.
An even simpler, minimal model for bacterial turbulence has been introduced in refs. \cite{wensink2012meso,dunkel2013minimal}. We here further investigate this model in two dimensions, in which the locally coarse-grained bacterial velocity field is considered as the only order parameter.
This assumption is based on the premise that in a dense suspension the local orientation of bacteria aligns with that of the velocity field.
The equations for the coarse-grained order parameter field $\bu$ take the form
\begin{eqnarray}
			\partial_t \bu + \lambda_0\bu\cdot\nabla\bu &=& -\nabla p - (\Gamma_0\Delta
				+\Gamma_2 \Delta^2+\alpha + \beta \bu^2)\bu \nonumber \\  
			 \nabla \cdot\bu&=&0
\label{eq:active}
\end{eqnarray}
The pressure gradient $\nabla p$ is the Lagrange multiplier ensuring incompressibility of the velocity field. The assumption of incompressibility is valid for dense suspensions. The free parameters $\lambda_0, \Gamma_0, \Gamma_2, \alpha$ and $\beta$ can be chosen to match experimental results \cite{wensink2012meso}. The parameter $\lambda_0$ is related to the type of the bacteria, i.e. whether they are of pusher or puller type. For pusher bacteria like {\it B. Subtilis}, $\lambda_0>1$. As discussed below, the number of parameters can be reduced by non-dimensionalizing the equations.
The linear terms in the above equation select a range of scales that are excited to model the forcing in the bacterial flow, which occurs predominantly at small scales. In Fourier space, the linear part of the equation can be written as $\gamma(k)\tilde{\bu}({\bs k}, t):=(\Gamma_0k^2-\Gamma_2k^4-\alpha)\tilde{\bu}({\bs k}, t)$. Consequently, the excited modes correspond to the ones where $\gamma(k) > 0$. The nonlinear advective term, like in the Navier-Stokes equation, is responsible for the energy transfer and thus allows for the formation of large-scale structures. The cubic term is a nonlinear saturation which together with the squared Laplacian term ensures the regularity of these equations \cite{zanger2015analysis}. A detailed description of these equations can be found in refs. \cite{dunkel2013minimal,dunkel2013fluid}. 

\begin{figure}
\resizebox{0.45\textwidth}{!}{
  \includegraphics{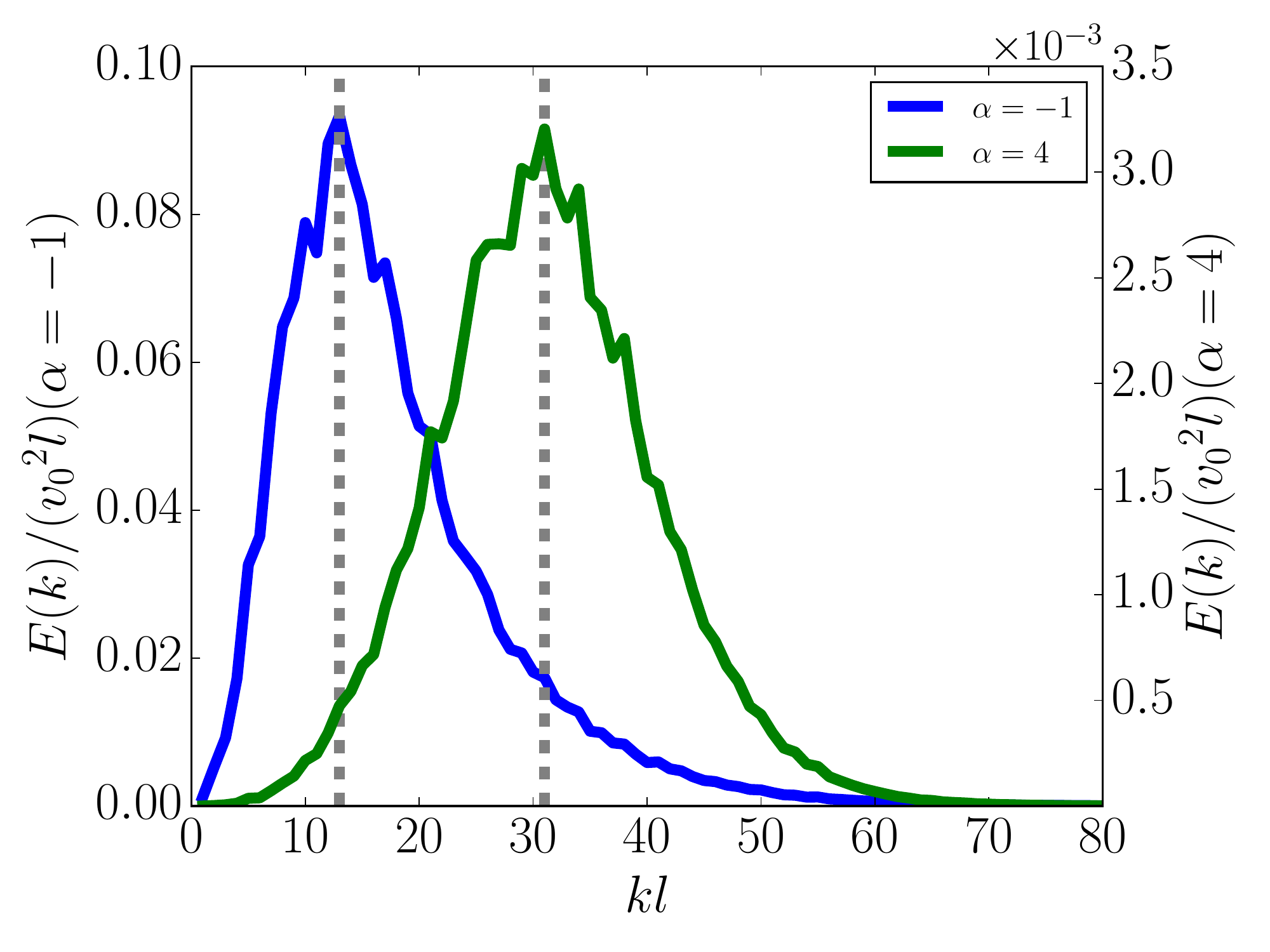}
}
\caption{Energy spectra of the velocity field for the active turbulence case (blue) and the weakly excited case (green). The dashed vertical lines indicate the wavenumber corresponding to the dominant scale in the system. In the active turbulence case, the energy spectrum peaks at a much larger length scale (lower wavenumber) due to the formation of meso-scale vortices in the system as a result of the inverse energy transfer.}
\label{fig:spectra}
\end{figure}

For the current investigations, we non-dimensionalize the equations following ref. \cite{bratanov2015new} and then normalize our numerical results based on dynamically emerging length and time scales. In summary, the procedure is as follows. The fastest growing linear mode $k_c=\sqrt{\Gamma_0/(2\Gamma_2)}$ is determined by the maximum of $\gamma(k)$. Consistent with ref. \cite{bratanov2015new} we select a length scale $l = 5\pi/k_c$. A velocity scale can be defined dimensionally as $v_0=\sqrt{\Gamma_0^3/\Gamma_2}$, which also selects a time-scale $l/v_0$. Non-dimensionalizing eq. \eqref{eq:active} using this length scale and time scale reduces the two parameters $\Gamma_0$ and $\Gamma_2$ to constant numbers $0.045$ and $9\times 10^{-5}$ respectively, thus decreasing the number of free parameters to three. If not noted otherwise, we choose the set of parameters $\lambda_0=3.5$, $\alpha=-1.0$ and $\beta=0.5$, which already has been investigated in \cite{bratanov2015new}. We normalize our numerical results with respect to the dominant length scale in the system. The wavenumber $k_{\mathrm{max}}$ corresponding to the peak of the energy spectrum (see fig. \ref{fig:spectra}) defines the dominant length scale in the system as $L = 2\pi/k_{\mathrm{max}}$.  This length scale can also be used to define a time scale given by $T = L/V$ where $V = \sqrt{\langle \bu^2\rangle}$. Such a procedure characterizes the significance of the dominant length scale in the system.

We numerically solve these equations in two dimensions by using a standard pseudospectral algorithm (with 1/2 dealiasing to account for the cubic nonlinearity) following a second-order Runge-Kutta scheme for time stepping with time step $0.0002$. We choose a domain size of $5\pi\times5\pi$ with $2048\times2048$ grid resolution.A large-scale flow is chosen as the initial condition. By testing different initial conditions, we ensured that the statistically stationary state is independent of the particular choices. For Lagrangian measurements, a million tracer particles are advected with the flow. The tracer particles are evolved according to the Lagrangian equations of motion $\mathrm{d} \bs X(\bs x_0,t) /  \mathrm{d} t = \bu(\bs X (\bs x_0,t),t)$, where $\bs X(\bs x_0,t)$ is the position of a tracer particle at time $t$ starting from ${\bs x_0}$ at time $t_0$. The velocity $\bu(\bs X(\bs x_0,t),t)$ at inter-grid points is interpolated by using a bicubic scheme. The system is allowed to evolve until it reaches a statistically stationary state after an approximate duration of $10T$ before measurements are taken. To identify and track vortex cores we follow an algorithm described in ref. \cite{giomi2015geometry}, details of which are given in sect. \ref{sec:lagrangian}.

\section{Eulerian statistics}
\label{sec:euler}
To connect to previous work \cite{wensink2012meso,bratanov2015new} as well as to set a reference point for the subsequent investigation of Lagrangian properties of the flow, we start with characterizing the Eulerian statistics of active turbulence.
Figure \ref{fig:eulerian}(b) shows the single-point velocity probability density function (PDF) of the active turbulence field. Since the flow is isotropic, we use one component of the velocity field to evaluate these PDFs. The distribution is close to Gaussian with sub-Gaussian tails. Sub-Gaussian tails for the single-point velocity have also been found for three-dimensional hydrodynamic turbulence \cite{wilczek11jfm}. The vorticity PDF, shown in fig. \ref{fig:eulerian}(c), departs strongly from Gaussianity with a comparably narrow core and wide tails, which roll off rapidly for large vorticity values. As is well known from the study of hydrodynamic turbulence, such departures from Gaussianity can be regarded as a signature of coherent vortex structures \cite{wilczek09pre,leung2012geometry,elsinga2010universal}. For example, they have also been observed in decaying two-dimensional Navier-Stokes turbulence \cite{wilczek08phd}.

\begin{figure}
\resizebox{0.45\textwidth}{!}{
  \includegraphics{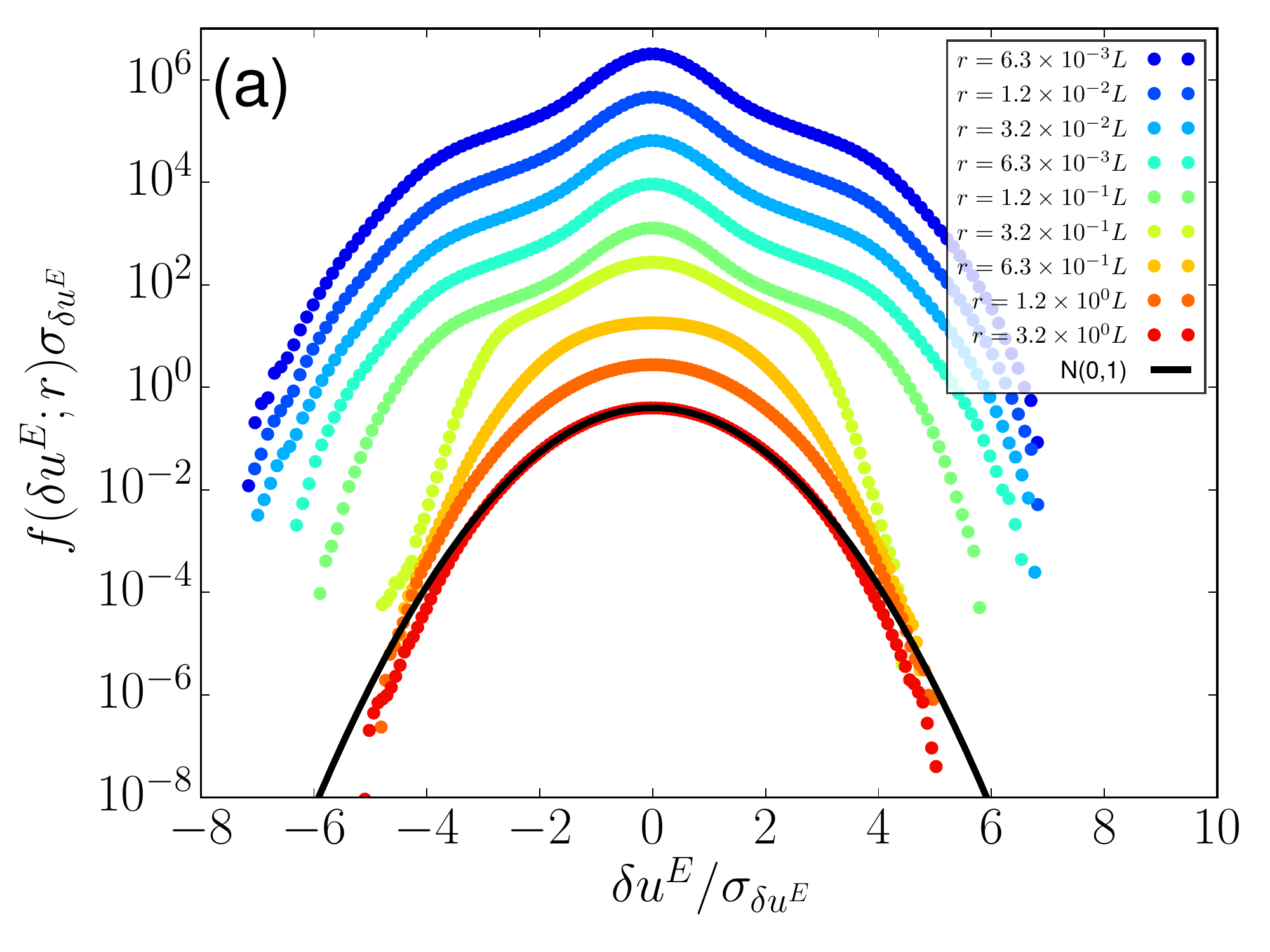}
}
\resizebox{0.45\textwidth}{!}{
  \includegraphics{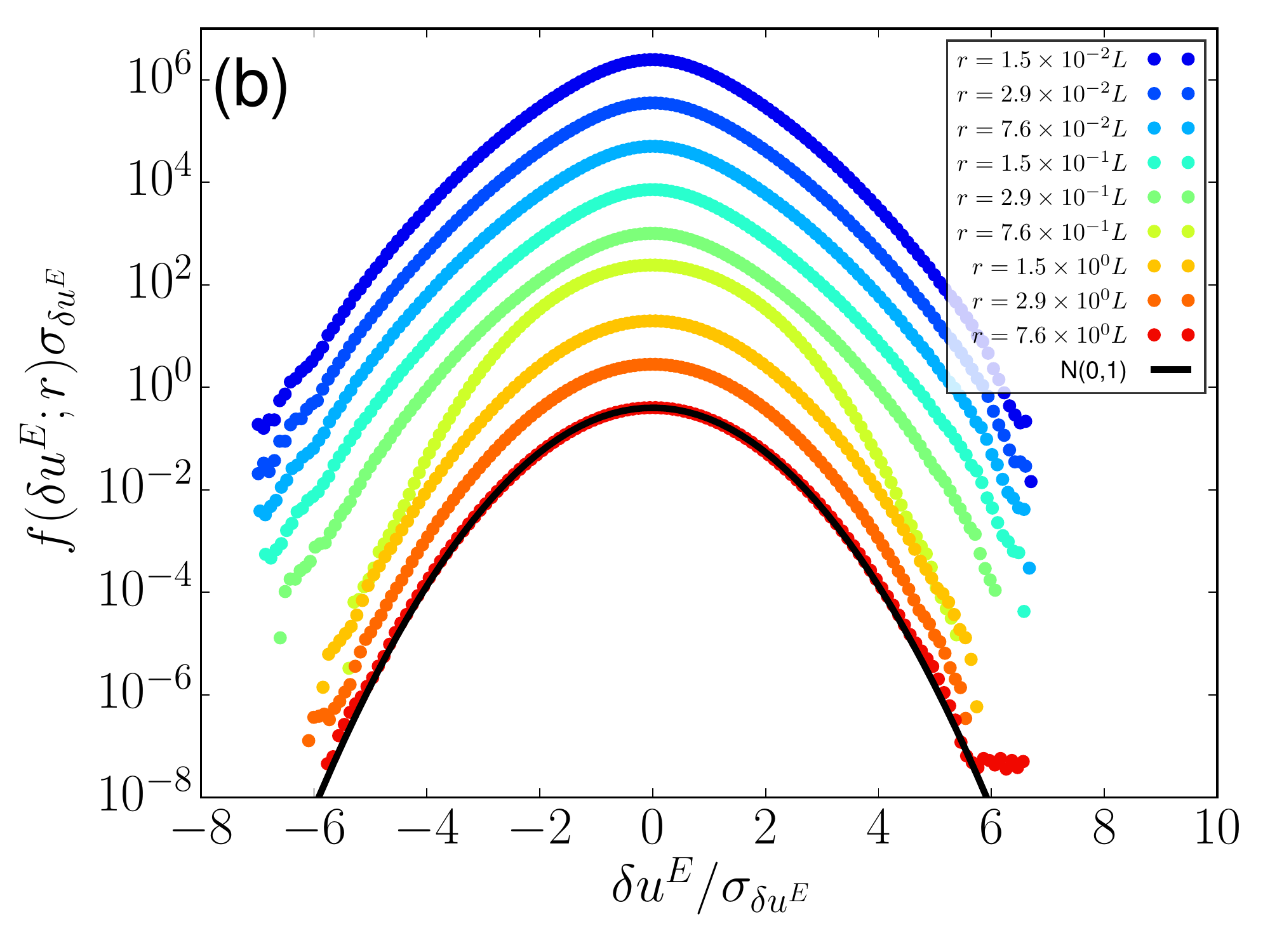}
}
\caption{Eulerian longitudinal velocity increment PDFs for (a) the active turbulence case and (b) the weakly excited case. The small-scale increment PDF for the active turbulent case displays considerable deviations from Gaussianity. In comparison, the weakly excited case with less pronounced vortex structures shows a close-to-Gaussian behavior at all scales.}
\label{fig:eulerianinc}
\end{figure}

The active turbulence model \eqref{eq:active} gives precise control over the energy injection mechanism, which motivated us to further investigate the influence of active forcing on non-Gaussian features of the flow. For the active turbulence case with $\alpha=-1$, the linear terms represented through $\gamma(k)$ introduce an active broadband forcing which predominantly injects energy at the wavenumber $k_c$. This broadband forcing can be reduced to a narrow band of wavenumbers with a reduced energy input by increasing the damping rate. Here, we consider the case with $\alpha=4$. The results of this numerical experiment are shown in the lower row of fig. \ref{fig:eulerian}. Figure \ref{fig:eulerian}(d) shows a snapshot of a vorticity field, which now displays less pronounced vortex structures compared to the active turbulence case. Still, the dynamics remains non-trivial as documented in the supplementary movie 2. The single-point velocity PDF shown in panel (e) is very close to Gaussian in the weakly excited case. Consistent with the observation of less pronounced vorticity structures, the vorticity PDF is now much closer to a Gaussian with slightly super-Gaussian tails.

To characterize multi-scale features of the flow, we obtain PDFs of the longitudinal velocity increments $\delta u^E=\left[\bs{u}(\bs{x}+\bs{r},t)-\bs{u}(\bs{x},t)\right]\cdot \bs{\hat{r}}$ for both cases as presented in fig. \ref{fig:eulerianinc}. Consistent with previously published results \cite{wensink2012meso,bratanov2015new} we find close-to-Gaussian PDFs from large to intermediate scales in the active turbulence case (panel (a)). Only at smaller scales on the order of $L$ we find departures. This change from Gaussian to non-Gaussian statistics occurs rather abruptly in scale, and can be accounted to the presence of meso-scale vortices in the flow. Consistent with this picture, the weakly excited case shows a close-to-Gaussian statistics for all considered cases (panel (b)).
\section{Vortex dynamics and Lagrangian transport properties}
\label{sec:lagrangian}
\begin{figure*}
\resizebox{0.328\textwidth}{!}{
 \includegraphics{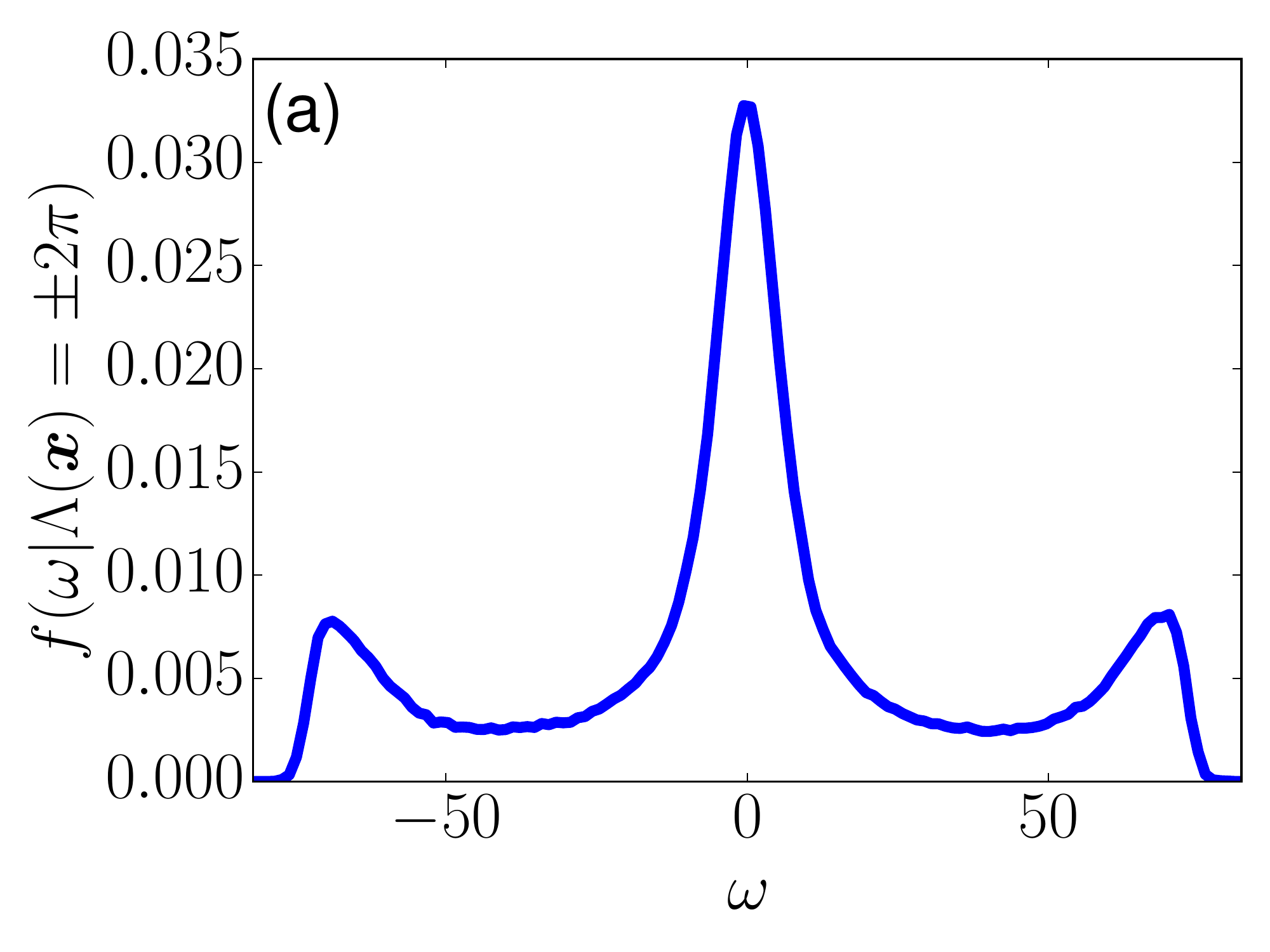}
}
\resizebox{0.328\textwidth}{!}{
 \includegraphics{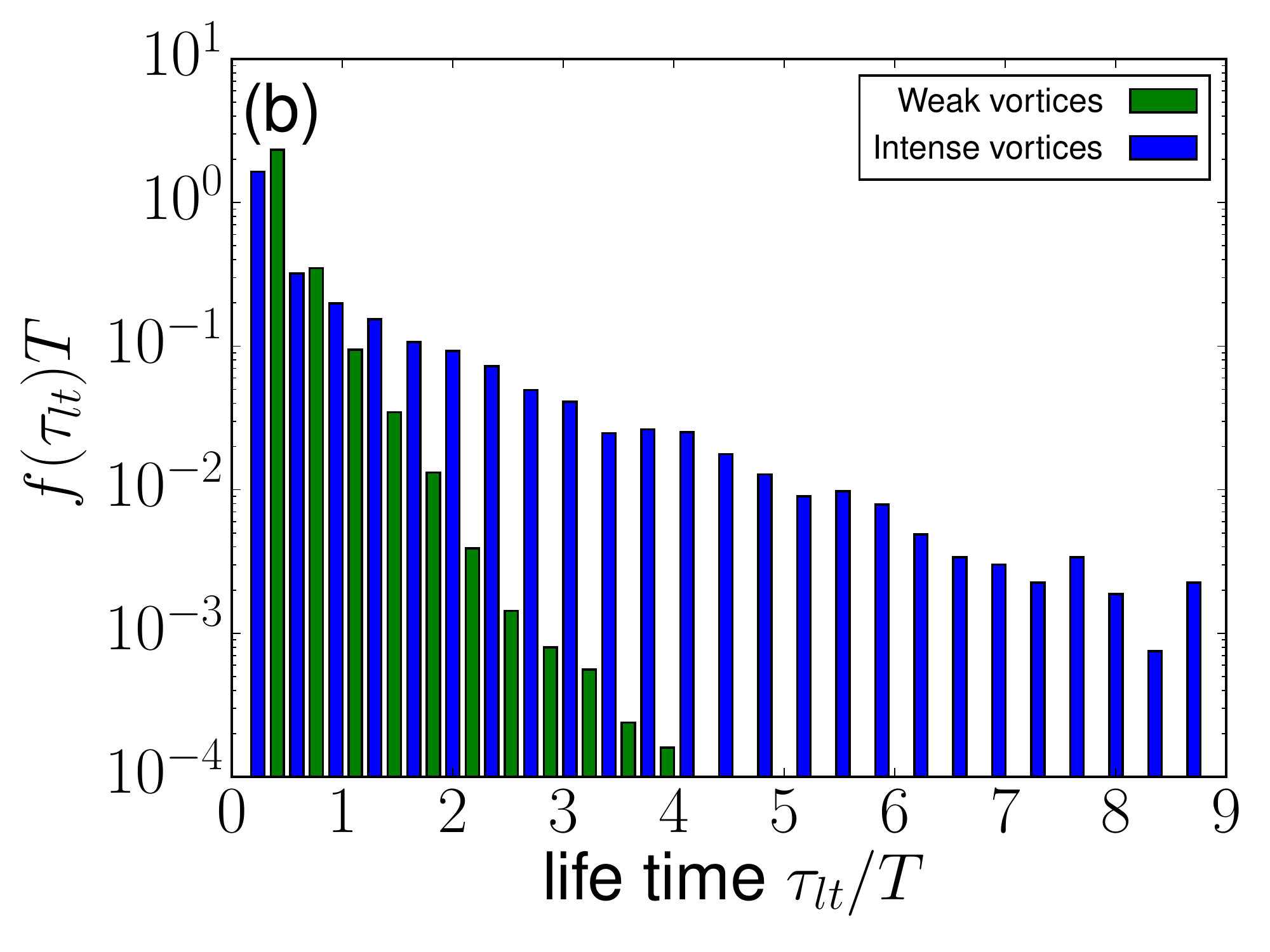}
}
\resizebox{0.328\textwidth}{!}{
 \includegraphics{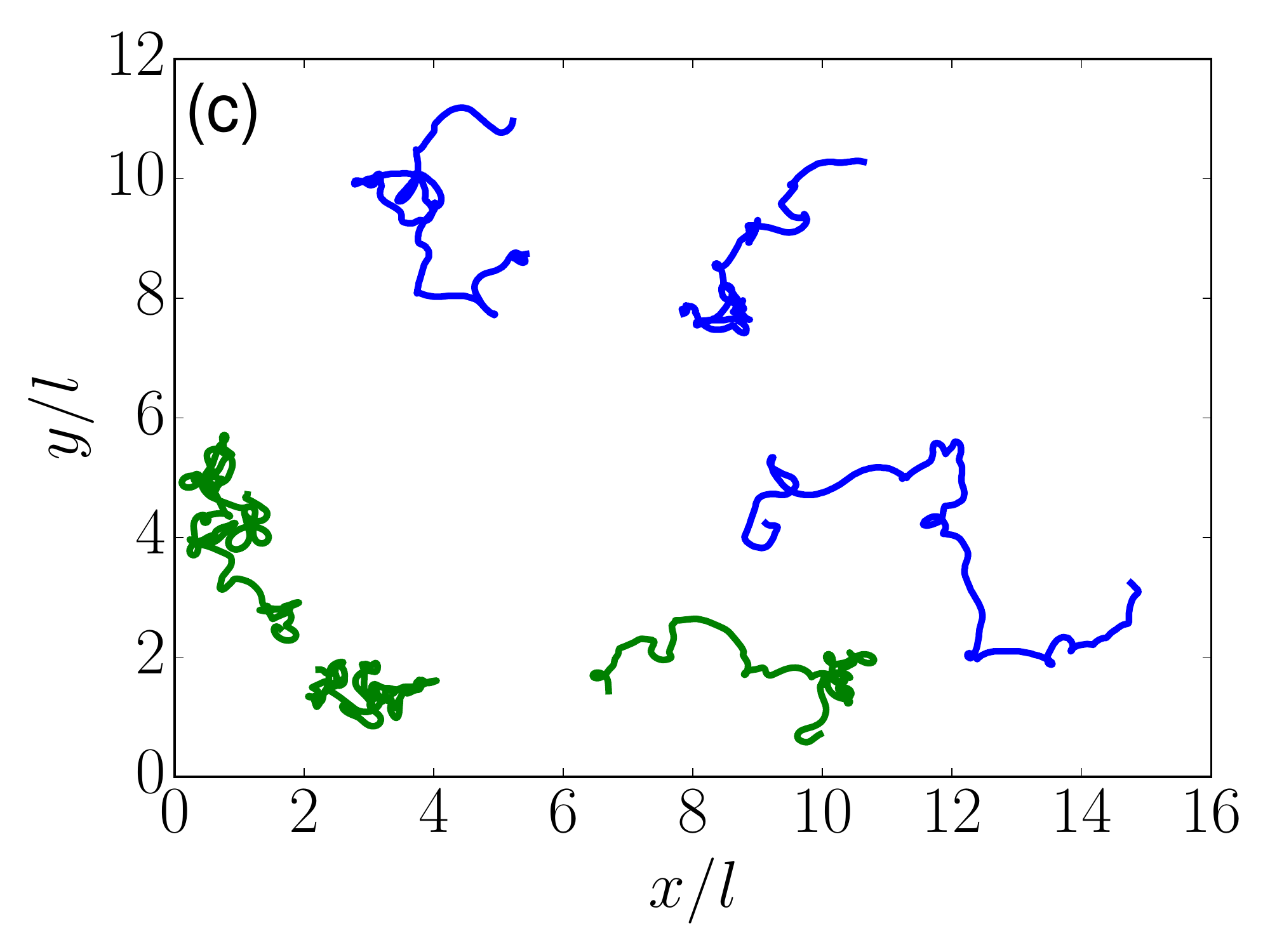}
}

 \resizebox{0.328\textwidth}{!}{
 \includegraphics{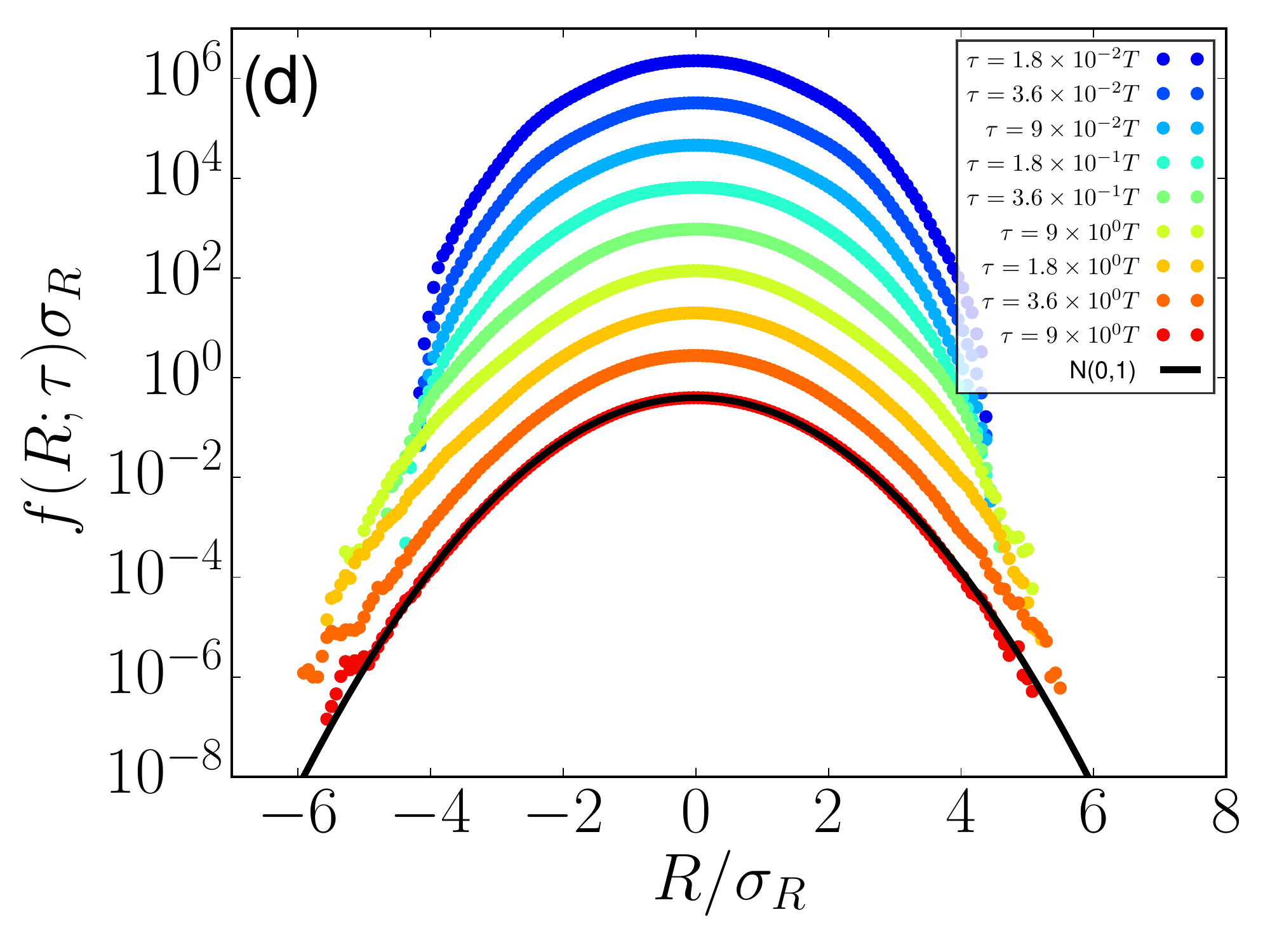}
}
\resizebox{0.328\textwidth}{!}{
 \includegraphics{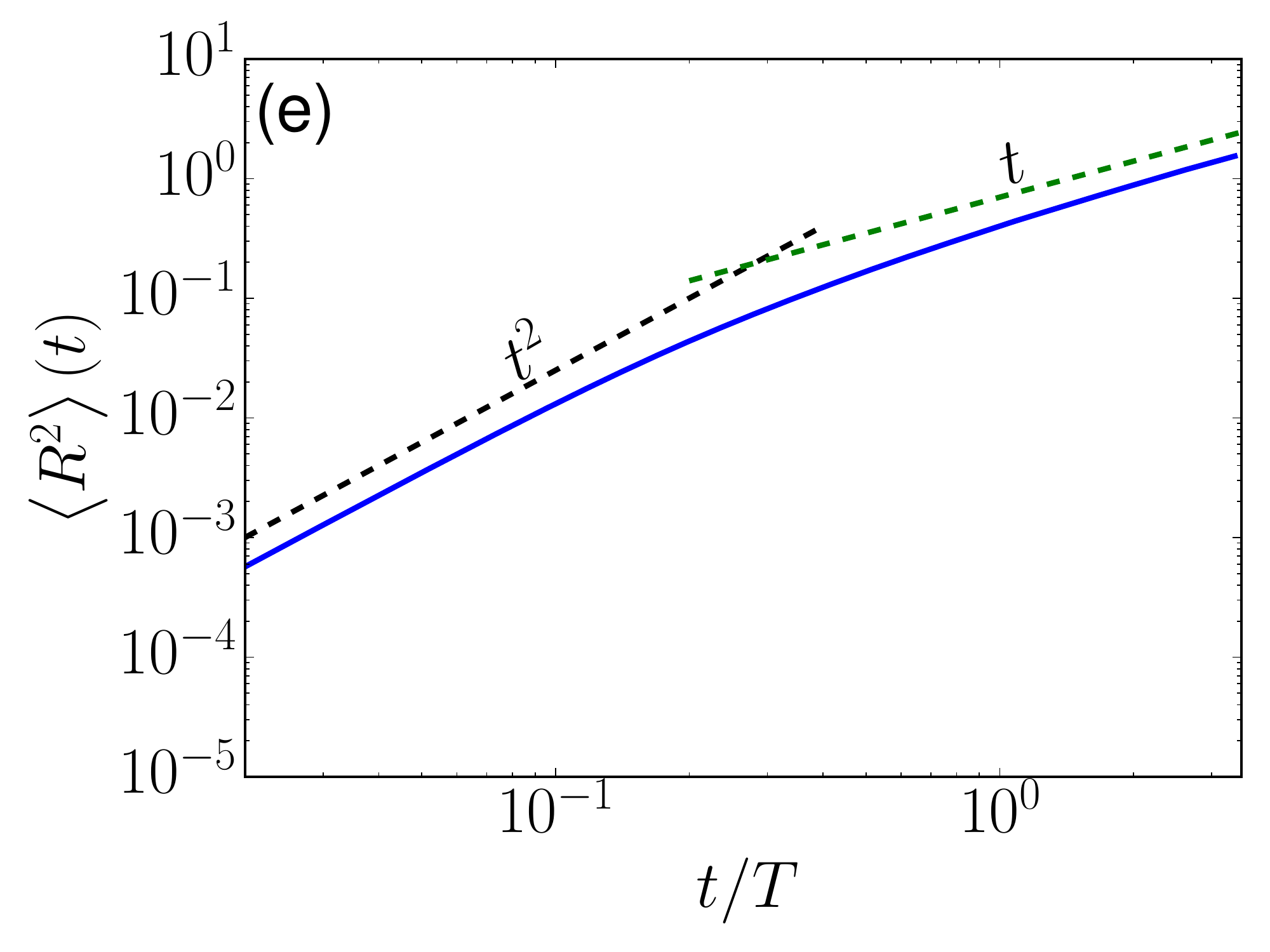}
}
 \resizebox{0.328\textwidth}{!}{
 \includegraphics{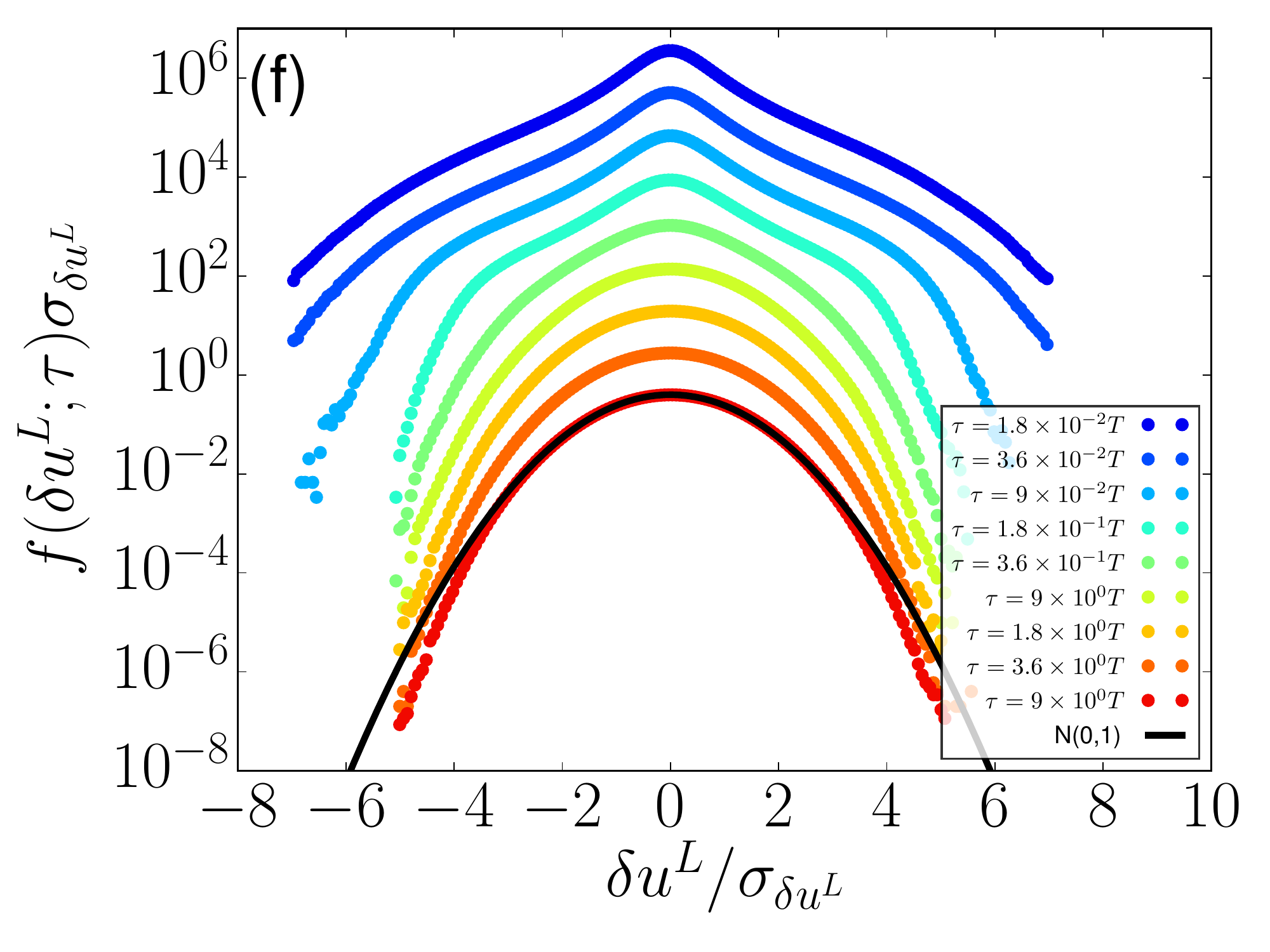}
 }
 \caption{Upper row: statistics and dynamics of active matter vortices. Panel (a) shows the distribution of vorticity at vortex cores. The different peaks correspond to the vortices of the two classes - weak and intense. The distribution of vortex lifetimes for weak and intense vortices is shown in panel (b). Note that the intense vortices have on average longer lifetimes. Some sample vortex core trajectories are shown in panel (c) (in blue) along with some passive Lagrangian tracer particles (in green).
 Lower row: Lagrangian statistics of active turbulence. Panel (d) shows the single-particle dispersion PDF which is close to Gaussian for all time lags considered. The mean squared displacement, shown in panel (e), exhibits a cross-over from a ballistic to a diffusive regime. Lagrangian velocity increment distributions are shown in panel (f).}
 \label{fig:vortexandlagrangian}
\end{figure*}

The results of the previous section have pointed out the significance of vortex structures, which is further investigated here, focusing on the vortex strength, dynamics and lifetimes. Some of these aspects are closely related to Lagrangian features of the flow, which are also analyzed in the following, where we restrict ourselves to the active turbulence case.
In the previous section, we analyzed single-time statistical features of the active turbulence field from an Eulerian point of view. To further characterize the role of vortex structures, vortex cores are identified as the centers of the cells around which the velocity vector takes a full rotation \cite{giomi2015geometry}. To this end, we calculate the angle $\Lambda$ which the velocity vector rotates around center point $\bs{x}$ of each cell, and vortex cores are defined as the centers of those cells where $\Lambda(\bs{x})=\pm2\pi$. This allows us to calculate vortex cores from the velocity field, although only with an accuracy of the grid resolution. Having identified the vortex cores, we are able to investigate their typical strength. Figure \ref{fig:vortexandlagrangian}(a) shows the distribution of vorticity at vortex cores. From the distribution it is clear that there are predominantly two classes - weak and intense vortices. The intense vortices correspond to the ones clearly identifiable in fig. \ref{fig:eulerian}(a) whereas the weak ones correspond to less coherent, not necessarily axisymmetric vorticity patches. The observation of the distinct large-amplitude vortices explains the sharp roll-off of the PDF shown in fig. \ref{fig:eulerian}(c). Assuming a typical profile for vortex structures, the vorticity PDF can be thought of as a smeared-out version of the vortex-core strength PDF.

The distribution of vortex core lifetimes, shown in fig. \ref{fig:vortexandlagrangian}(b), further clarifies the difference between weak and intense vortices. As can be inferred from the PDF, weak vortices typically have a much shorter lifetime than intense vortices. Also, the lifetime statistics for both weak and intense vortices decay to zero rather rapidly. We note that there are events in which an intense vortex transforms into a weak vortex and vice versa, which is one limiting factor of their lifetimes. Figure \ref{fig:vortexandlagrangian}(c) shows some trajectories of vortices with long lifetimes along with some typical Lagrangian tracer trajectories. In comparison, the vortices appear to have smoother trajectories than the tracer particles. This is because the vortex core trajectories by design pick out very specific points in the flow field. In contrast to vortex cores, typical Lagrangian tracer particles encounter a number of vortex trapping events, which explains their rapid coiling. Similar observations have been made in two-dimensional turbulence \cite{wilczek08phd,kamps08pre}.
 
Proceeding to the characterization of Lagrangian statistics, fig. \ref{fig:vortexandlagrangian}(d) shows standardized PDFs of Lagrangian single-particle dispersion, where $R$ is one component of ${\boldsymbol X}({\boldsymbol x_0}, t_0+\tau)-{\boldsymbol x_0}$. Owing to the approximate Gaussianity of the velocity, the PDFs are close to Gaussian for short time scales, and the Gaussianity persists up to the largest time scales, similar to what is found in hydrodynamic turbulence. The mean squared displacement, which characterizes the variance of this approximately self-similar process in scale, is shown in fig. \ref{fig:vortexandlagrangian}(e). As expected, the displacement is initially ballistic, i.e. it scales as $t^2$, and then transitions to a diffusive long-time behavior with a scaling proportional to $t$. The transition occurs on the order of the time scale $T$ which characterizes approximately the time spent by a tracer particle in a vortex.

Finally, we wish to characterize temporal velocity fluctuations along Lagrangian tracer particles. Figure \ref{fig:vortexandlagrangian}(f) shows the distribution of Lagrangian velocity increments $\delta u^L$, defined as either of the components of ${\boldsymbol u}({\boldsymbol X}({\boldsymbol x_0}, t_0+\tau), t_0+\tau)-{\boldsymbol u}({\boldsymbol x_0},t_0)$, for different values of time lag $\tau$. We observe that for short time scales, the statistics of the Lagrangian velocity increment shows strong deviations from Gaussianity, consistent with the observation for the Eulerian increments. In the limit of $\tau \rightarrow 0$ the velocity increment is proportional to the single-particle acceleration. Like in the Eulerian frame, the PDFs relax to a Gaussian shape rather sharply at a value of $\tau\approx T$, strengthening the hypothesis that the velocity field in active turbulence has a simple structure beyond the length scale of the individual strong vortices.
\section{Summary and Conclusions}
\label{sec:conclusions}

In this work we have conducted a statistical study of a minimal continuum model for bacterial turbulence. The numerical and statistical results show that active turbulence displays close-to-Gaussian statistics both in an Eulerian and a Lagrangian frame when moderate to large scales are considered. Similar observations can be made in statistically stationary two-dimensional turbulence with large-scale friction. Deviations are found at the scale where coherent vortices occur, as can be probed with vorticity and velocity increment statistics. Employing a vortex identification and tracking algorithm, we find that active turbulence selects intense vortices of rather well-defined magnitude, which is in contrast to hydrodynamic turbulence. The life-time statistics of the vortices are investigated and display a rapid fall-off to zero. By increasing the damping rate compared to the active turbulence case, the emergence of meso-scale vortices is found to be suppressed, which goes along with statistics even closer to Gaussianity, corroborating the connection between non-Gaussian statistics and vortex structures.

The meso- to large-scale Gaussianity of active turbulence may open avenues for future analytical modeling approaches. A natural next step in the direction of this work is the development of a statistical theory of active turbulence, which is the subject of ongoing research.

\vspace*{0.1in}
This work was supported by the Max Planck society. MJ gratefully acknowledges the financial support by the International Max Planck Research School ``Physics of Biological and Complex Systems'', G\"{o}ttingen.

The final publication is available at Springer via http://dx.doi.org/10.1140/epje/i2018-11625-8
\section*{Authors contribution statement}
M.W. conceived the study. M.J. performed the simulations. Both authors analyzed the results and wrote the manuscript.
\bibliography{stat_active}
\bibliographystyle{unsrt}
\end{document}